\documentstyle[aps,eqsecnum,epsf,floats]{revtex} 

\begin{document}
\draft
\twocolumn[\hsize\textwidth\columnwidth\hsize\csname
           @twocolumnfalse\endcsname

\title{Truncated post-Newtonian Neutron Star Model}
\author{Hisa-aki Shinkai  \cite{Email-his}}
\address{ 
Department of Physics, Washington University, 
St. Louis, Missouri 63130-4899, USA
}
\date{May 15, 1998 / revised June 27, 1998 / accepted for publication
in PRD. (Brief Report)  \cite{status}}
\maketitle
\begin{abstract}
\widetext

As a preliminary step towards simulating binary neutron star
coalescing problem, we test a post-Newtonian approach by constructing
a single neutron star model. 
We expand the 
Tolman-Oppenheimer-Volkov equation of hydrostatic
equilibrium by the power of $c^{-2}$, where $c$ is the speed of light,
and truncate at the various order.  We solve the system 
using the polytropic equation of state with index $\Gamma=5/3, 2$ and 3,
and show how this approximation converges together with mass-radius relations. 
Next, we solve the Hamiltonian constraint equation with 
these density profiles as trial functions, and examine the differences
in the final metric.  
We conclude the second `post-Newtonian' approximation
is close enough to describe general relativistic single star.   
The result of this report will be useful for 
further binary studies. 
\end{abstract}

\vskip 2pc]
\narrowtext


\section{Introduction} \label{sec:intro}
Several earth-based interferometers designed to detect 
gravitational waves have been recently constructed.  Detectors 
such as LIGO, VIRGO, GEO and TAMA 
are expected to begin operating
within a few years (see {\it e.g.} \cite{gw-review}).
In order to extract gravitational waveforms from noisy data
and to discuss physical parameters, it is essential to predict
waveforms in advance by both analytical and numerical approaches. 

Binary neutron star systems are one of the most plausible sources
of gravitational waves. 
They emit energy through gravitational radiation,  
shrink their inspiral orbits gradually, and finally merge with 
strong emission of gravitational waves. 
The system is described by the post-Newtonian (PN) approximation 
(see {\it e.g.} \cite{will94}) 
in the last several minutes before they merge, while
in the last phase of coalescence of stars
we need to solve the Einstein equations which are available
only through numerical integration. 

After the pioneering numerical works by Nakamura and Oohara in the 
Newtonian gravity with radiation reaction correction \cite{kyoto-newt}, 
several groups started developing their numerical codes to solve this
problem in a more realistic way. 
Such hydrodynamical simulations are categorized as in the 
Newtonian scheme (with/without radiation reaction term)
\cite{kyoto-newt2,cornell-newt,drexel-newt,piran-newt,max-newt,new-tohline,swesty,uryu};
PN approximation
\cite{kyoto-pn}; 
and fully general relativistic level
\cite{wilson,kyoto-gr,cactus2}. 
However, we do not have a method to construct 
physically satisfactory initial data for inspiral
binaries in general relativity.
Most of the numerical tests
start their simulations under 
assumptions of certain quasi-equilibrium and   
conformal flatness of spacetime, with a particular choice of 
vorticity of fluid ({\it e.g,} \cite{BGM97} and references therein). 

One way to prepare initial data might be patching the
PN scheme to the general relativistic one \cite{mg8}. 
In this report, 
we construct a simple model and examine how this effort is justified. 
We solve the Tolman-Oppenheimer-Volkov (TOV) equation of hydrostatic
equilibrium of a single neutron star, which is truncated at the various
PN levels. 
We compare the  mass and radius of a star
as a function of central density
using the polytropic equation of state.
We also solve the  Hamiltonian constraint equation of the Einstein
equations by substituting these density profiles as trial functions, and 
discuss the differences in the metric. 

This study is an extended one from the earlier works
\cite{wagoner,ciufolini,castagnino,lombardi}
 at the first PN approximation. 
We intend to make a bridge between the Newtonian and general 
relativistic solutions
of a neutron star model, both of which are first shown numerically by 
Tooper\cite{tooper}. 

In the actual calculations, we used the geometrical units of 
$c=G=M_\odot=1$, where $c, G, M_\odot$ are the speed of light, 
Newton's gravitational constant and the solar mass, respectively.  However, 
$c$ and $G$ will appear in the text where they help understanding.

\section{Truncated TOV neutron stars}\label{sec:tovpn}
In general relativity (GR), we have the TOV equation for solving
a hydrostatic equilibrium star in the spherically symmetric spacetime. 
We start from the metric
\begin{equation}
ds^2 = - e^{2\Phi(r)} dt^2 + e^{2\Lambda(r)} dr^2 + r^2\left( d\theta^2
+ \sin^2 \theta d\varphi^2 \right), \label{metric_tov1}
\end{equation}
where 
$e^{2\Lambda(r)} = (1-{2Gm(r) \over c^2 r})^{-1}$.
Then the TOV equations are written as 
\begin{eqnarray}
{dm \over dr}&=& 4 \pi r^2 \rho_t, \\
{dp \over dr}&=& -{Gm \rho_t \over r^2} (1 + {p \over \rho_t c^2} )
 ( 1+{4 \pi p r^3 \over m c^2})(1-{2Gm \over r c^2})^{-1}, \label{TOV2}\\
{d\Phi \over dr}&=& -{1 \over \rho_t}
{dp \over dr}(1+{p \over \rho_t c^2})^{-1},
\label{TOV3}
\end{eqnarray}
together with the specified equation of state, for which we use the 
polytropic equation of state
\begin{equation} p=K \rho^\Gamma = K \rho^{1+1/n}, \end{equation}
where $p$, $\rho$ are the pressure and energy density, respectively, and 
 $\rho_t$  is the total mass density,  
\begin{equation}
\rho_t= 
\rho + {p \over (\Gamma -1) c^2}.
\end{equation}
Obviously, the set of equations recover the Newtonian limit for
$c^2 \rightarrow \infty$.

The idea of this report is to 
expand the product of the
parentheses in (\ref{TOV2}) and (\ref{TOV3}) and truncate them 
at the order of $1/c^{2i}$. The $i$-th truncation, then, gives the
so-called $i$-th PN approximation.
(The case of $i=1$ is briefly mentioned in \cite{KWbook}.)
That is, we write 
 (\ref{TOV2}) and (\ref{TOV3}) schematically
\begin{eqnarray}
{dp \over dr}&=&-{Gm \rho_t \over r^2} (1+A)(1+B)(1-C)^{-1} \nonumber \\
&=&-{Gm \rho_t \over r^2} (1+ A+B+C \nonumber \\
&& ~~~~~  +AB+AC+BC+C^2+\cdots) \label{pNexp1}  \\
{d\Phi \over dr}&=& -{1 \over \rho_t}{dp \over dr}(1+A)^{-1} \nonumber \\
&=& -{1 \over \rho_t}{dp \over dr}(1-A+A^2-A^3+\cdots).\label{pNexp2}
\end{eqnarray}
If we use 
 these equations with terms in the RHS of up to two products of $A, B, C$ 
(such as $AB$ or $A^2$), then we say the system is in the second 
PN approximation.

\begin{figure}[tbp]
\setlength{\unitlength}{1in}
\begin{picture}(3.75,6.6)
\put(-0.2,0.0){\epsfxsize=4.25in \epsfysize=6.36in 
              \epsffile{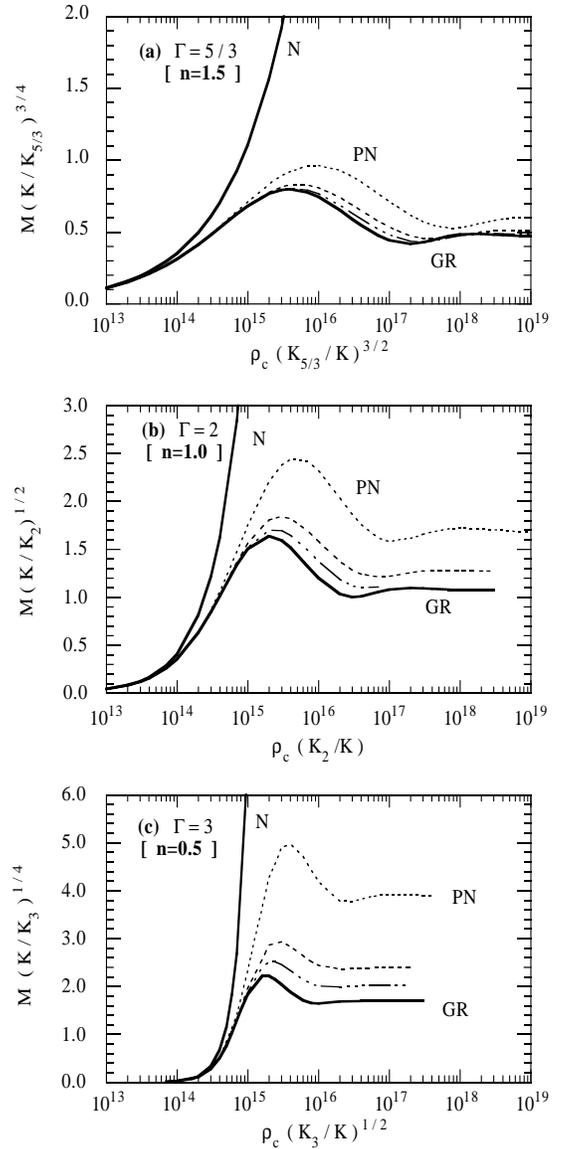} }
\end{picture}
\caption[fig1]{
Total mass as the function of its central density 
for truncated neutron star model. 
Figures (a), (b) and (c) are for different equation of state
with $\Gamma=5/3$, 2 and 3, respectively. 
Mass is in the unit of solar mass and central density is in
[g/cm$^3$]. 
The gray solid line is of Newtonian solutions, the solid line 
is of general relativistic solution. 
The 
dotted line, dashed line and three-dot-line
are of first, second and third post-Newtonian
approximated solution, respectively.}
\label{fig1}
\end{figure}

We apply $\Gamma=5/3, 2$
and 3 for the equation of state ($n=1.5, 1$ and 0.5 in the polytropic
index, respectively) and compare the solutions of Newtonian, GR
and up to third PN approximation. 

The radius of the star, $R$, is measured at the point, $r_\star$,
where  
density $\rho_t$ drops low enough [$O(10^{-10})$ in the 
geometrical units], and given by the proper 
length,
\begin{equation}
R=\int_0^{r_\star} \left(1 - {2 G m(r) \over c^2 r } \right)^{-1/2} dr,
\end{equation}
with appropriate truncation in the integrand. We express
the mass of the star, $M$, by $M=m(r_\star)$. 

We use 5th order Runge-Kutta method (Fehlberg method) to integrate 
the equations. In order to check that this approach is right, we also
worked the TOV equations in the harmonic gauge and confirmed
that we get the identical physical quantities in the results.

In Fig.\ref{fig1}, we show 
the total mass $M$ as the function of the 
 central density $\rho_c$ for the 
different  $\Gamma$s and PN levels. 
Mass is in the unit of $M_\odot$ and central density is in
[g/cm$^3$], and both are rescalable with the constant $K$
 in the equation
of state.  Here we use  $K$ in the calculations as:
 $K_{5/3}=4.35$  (for $\Gamma=5/3$),
 $K_{2}=10^2$ (for $\Gamma=2$), and  
 $K_{3}=10^5$  (for $\Gamma=3$) in the geometrical unit, 
where $K_{5/3}$ is the number for the pure neutron 
equation of state \cite{STbook}.

We see clearly the convergence of this PN approximation 
in all the $\Gamma$s. 
However, if the equation of state is stiff, then the high density
configuration differs from that of GR even at the 
higher PN approximation. 

From the first PN approximation, we see the existence
of the maximum mass. 
The central density which gives this maximum becomes larger in the
weak gravity approximation.

\begin{figure}[tbp]
\setlength{\unitlength}{1in}
\begin{picture}(3.75,6.6)
\put(-0.2,0.0){\epsfxsize=4.25in \epsfysize=6.36in 
              \epsffile{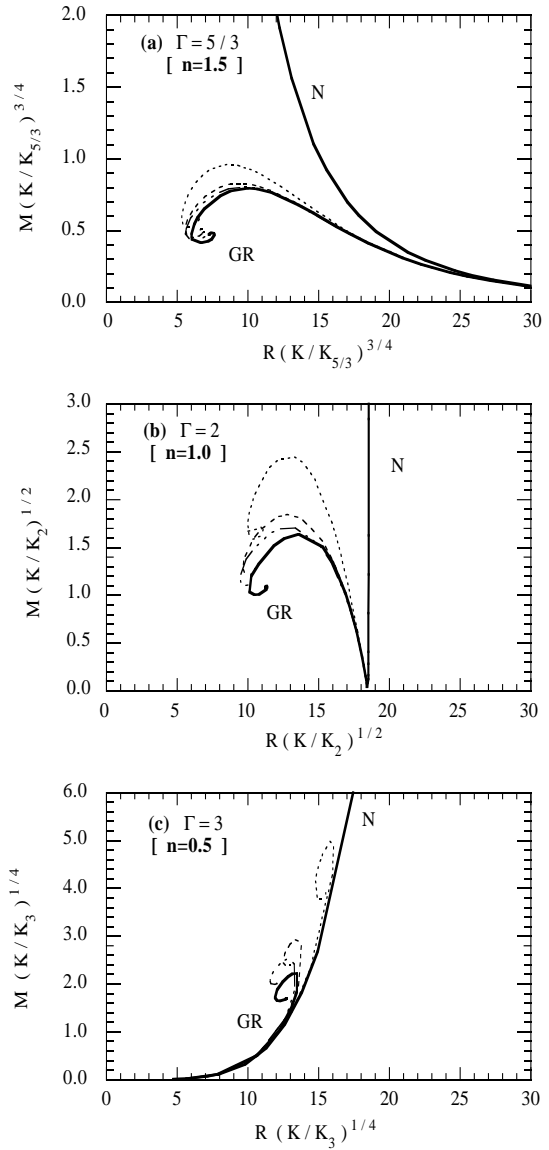} }
\end{picture}
\caption[fig2]{
Mass and radius relations for truncated neutron star models.
Mass is in the unit of solar mass and radius is in
[km]. 
The lines are the same as of  Fig.\ref{fig1}.}
\label{fig2}
\end{figure}

In Fig.\ref{fig2}, we show the 
mass-radius relations. 
In the Newtonian limit, the asymptotic behaviors of $M$
near $M=0$ are as 
$M \propto R^{-3}$ (for $\Gamma=5/3$), 
$M \propto R^{0}$ (for $\Gamma=2$) and
$M \propto R^{5}$  (for $\Gamma=3$). These represent
softness  (for $\Gamma=5/3$) and stiffness  (for $\Gamma=3$)
of the equation of state. We see that all the lines in Fig.\ref{fig2}
coincide with this Newtonian limit in the lower mass limit. 
The figure also shows us that the first PN solution has the
same feature as GR. 

We also checked the causality constraint $dp/d\rho \leq 1$ 
(see {\it e.g.} \cite{geroch_lindblom} ) in all of the models,
and confirmed that the constraint is always valid.

\section{Metric Output via Hamiltonian constraint} \label{sec:metric}

We next solve the Hamiltonian constraint equation in GR
with the trial density profiles obtained above. 
Our aim is to compare the difference of the output metric 
and to examine a matching scheme of PN data to the general
relativistic one. 

We use O'Murchadha-York's conformal approach \cite{york} to solve the 
Hamiltonian constraint. 
Defining the conformal factor $\psi$ and setting
$\gamma_{ij} = \psi^4 \hat{\gamma}_{ij}$, 
the constraint becomes
\begin{equation}
~8~^{(3)\!}\hat{\Delta}
\psi = ~^{(3)\!}\hat{R}\psi
- 16 \pi G \hat{\rho}\psi^{-3}
\label{inithamilt} 
\end{equation}
where $^{(3)\!}\hat{\Delta}$ and $~^{(3)\!}\hat{R}$
are the 3-dimensional Laplacian and Ricci scalar curvature, respectively, 
defined by $\hat{\gamma}_{ij}$.
Here we assumed $K_{ij}=\hat{K}_{ij}=0$. 

We choose our trial metric 
 $\hat{\gamma}_{ij}$ as conformally flat, and solve (\ref{inithamilt})
with a trial density configurations of 
$\hat{\rho}=
\rho_t$. 
We use the Incomplete Cholesky conjugate gradient (ICCG) method \cite{ICCG}
with the Robin boundary condition
$\psi= 1 + {C / r} $, where $C$ is a constant, 
 for solving  (\ref{inithamilt}). 

\begin{figure}[tbp]
\setlength{\unitlength}{1in}
\begin{picture}(3.75,6.6)
\put(-0.3,0.0){\epsfxsize=4.25in \epsfysize=6.36in 
              \epsffile{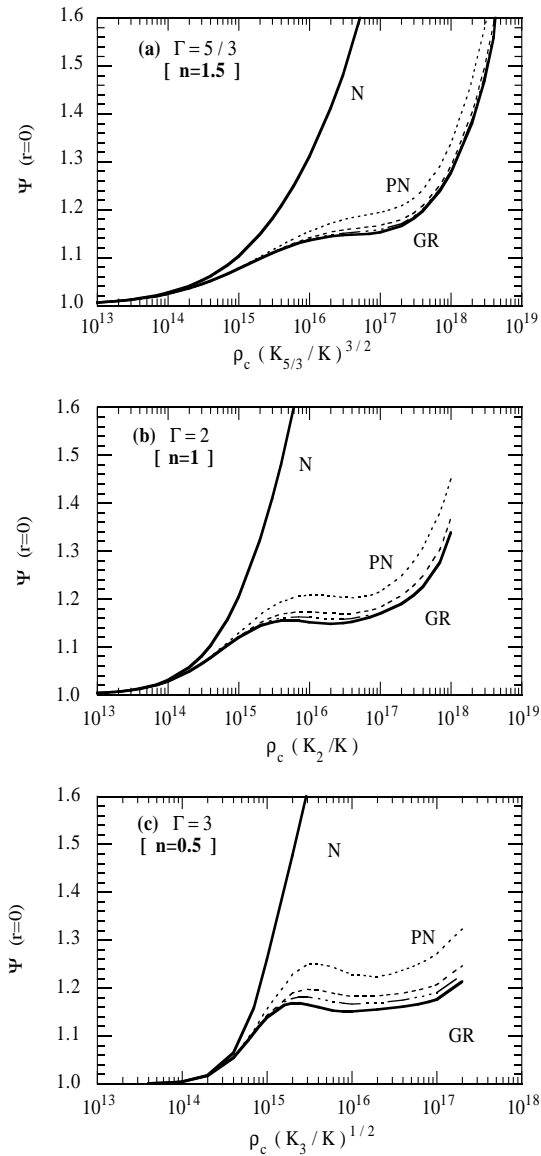} }
\end{picture}
\caption[fig3]{
The conformal factor $\psi$ at the origin 
is displayed as a function of 
central density, of which 
we used a trial configuration for solving Hamiltonian constraint
equation. 
The central density is in the unit of [g/cm$^3$].
Each line indicates the trial profile as input, using the same
notation with  Fig.\ref{fig1}.}
\label{fig3}
\end{figure}

In Fig.\ref{fig3}, we show the conformal factor $\psi$ at the origin 
as a function of central density of trial configuration. 
The 3-metric at the center will be given by 
$\gamma_{ij} = \psi^4 {\delta}_{ij}$. 
We see that using the Newotnian configuration as input gives us quite
different solutions from the expected ones of GR,   
while all PN trials give similar solutions with GR.  
Independently to $\Gamma$, we can say second PN approximation 
provides closer values for the output metric to those of GR.


\section{Discussion} \label{sec:disc}

In order to justify the recent post-Newtonian (PN) approaches to the 
binary
neutron star problem, we constructed a simple model.  By solving the 
hydrostatic equilibrium equation of a star at $i$-th PN 
approximation, we showed the convergence of this approach, the mass and
radius relations and resultant metric output via the Hamiltonian
constraint equation. 

We conclude that 
second  PN approximation 
provides quite similar density profiles
to those of GR, 
independent of equations of states. 
If we use second PN density configurations as trial functions, 
we get closer metric 
solutions to those from GR through the Hamiltonian constraint.  
Although this study is restricted to a hydrostatic single star model,
we think that the figures shown here are convenient templates for 
further numerical studies. 

As shown in \cite{mg8}, the discontinuous matching surface of PN and GR 
in the vacuum region will be smoothed out in fully relativistic 
evolution in a 
particular slicing condition.  Therefore we expect that 
higher PN initial data will smoothly evolve in the fully 
relativistic simulations, although there are many unknown factors as to
whether such an  initial data is numerically satisfactory or not. 
We are now applying this approach to construct a binary model 
including their velocity corrections together with fully general 
relativistic
hydrodynamical evolutions. 
This effort will be reported elsewhere.



\noindent
{\bf Acknowledgments}  ~~
The author thanks Stephen B. Selipsky, Wai-Mo Suen and Cliff. M. Will 
for discussions.  He also thank Ed Seidel and Doug Swesty for the
comments on the causality constraint. 
He appreciates the annonymous referee's suggestions in the section IV.
This work was partially supported by NSF PHYS 96-00049, 96-00507,
and NASA NCCS 5-153.



\begin{thebibliography}{99}
\bibitem[*]{status} 
This paper was accepted for publication in Physical Review D. [access
code dsj637]. However, since I was strongly suggested that the
contents of this paper should be included as a section in our group's
future paper, I gave up the publication.
\end{thebibliography}
\end{document}